\documentclass[preprint,showpacs,12pt,aps,bm,epsfig,amssymb]{revtex4}
\usepackage{epsfig,amsmath,amsfonts,float}
\newcommand{\R}[0]{{\mathbb{R}}}

\def\R{{\mathbb{R}}}

\newcommand{\bef}{\begin{figure}}
\newcommand{\eef}{\end{figure}}

\newcommand{\leb}{\left(}
\newcommand{\rib}{\right)}
\newcommand{\bei}{\begin{itemize}}
\newcommand{\eei}{\end{itemize}}
\newcommand{\bea}{\begin{eqnarray}}

\newcommand{\eea}{\end{eqnarray}}
\newcommand{\bequ}{\begin{equation}}
\newcommand{\eequ}{\end{equation}}

\DeclareRobustCommand\openone{\leavevmode\hbox{\small1\normalsize\kern-.33em1}}

\begin{document} 

\title{Entanglement generation via scattering of two particles with hard--core repulsion}
%
% Alternativ: Entanglement generation via scattering of two particles: a case study
% 
\author{Frank Schm\"user\footnote{\; electronic mail: schmues@ira.uka.de} and 
Dominik Janzing\footnote{\; electronic mail: janzing@ira.uka.de}}
\affiliation{Institut f\"ur Algorithmen und Kognitive Systeme, 
Universit\"at Karlsruhe, Am Fasanengarten 5, 76131 Karlsruhe, Germany.}
\date{Jan 26, 2006}

\begin{abstract}
We analyse the entanglement generation in a one dimensional scattering process. The two 
colliding particles have a Gaussian wave function and interact by hard--core repulsion.
In our analysis results on the entanglement of two mode Gaussian states are used. 
The  produced entanglement depends in a non-obvious way
on the parameters ratio of masses and initial widths.
The asymptotic wave function of the two particles and its associated 
ellipse yield additional geometric insight into these conditions. 
The difference to the quantitative 
analysis of the amount of entanglement generated by beam splitters with 
squeezed light is discussed.

%Analogies 
%of the scattering process with 
%situations in quantum optics are pointed out.
\end{abstract}
\maketitle
\section{Introduction}
In the last decade the study of entanglement has become one of the major topics in the flourishing
field of quantum information theory (for introductions into this topic see 
e.~g.~\cite{wootent,poprohr}). Among other things, it has 
become clear that entanglement is an essential resource for many desirable operations in quantum
information theory, for example quantum teleportation \cite{bentelep}
or some protocols in quantum cryptography \cite{qcekert}. 
For a detailed understanding of entanglement it is necessary 
to quantify it. For pure states $ | \psi \rangle \in {\cal H}_1 \otimes
{\cal H}_2 $ this is fairly straightforward. We consider the reduced density operator in $ {\cal H}_1 $
(tracing out system $2$) 
\bequ
\rho^1 := {\rm Tr}_2 \leb  | \psi \rangle \, \langle  \psi | \rib 
\eequ
and define the entropy of entanglement $ E \leb  | \psi \rangle \rib $  of the pure state $  | \psi \rangle $ 
as the von Neumann entropy of this reduced density operator 
\bequ
E \leb  | \psi \rangle \rib := S \leb \rho^1  \rib \quad {\rm with} \quad S \leb  \rho \rib := -
{\rm Tr} \leb \rho \; \log_2 \,  \rho \rib \quad .
\label{basedefi}
\eequ
The quantity $ E \leb | \psi \rangle \rib $ has the nice property that it (asymptotically) 
describes the 
number of maximally entangled singlet (or Bell) states of 2 qubits that can be obtained from
many copies of the quantum state $ | \psi \rangle $
when it is shared by two parties and each is allowed to operate on
his part only \cite{benconcentr}. The quantification 
of entanglement for mixed states is a much more difficult business which
however does not concern us in this paper, as we are only dealing with pure states. 

In this paper we study entanglement generation in 
quantum systems with a continuous degree of freedom, 
in particular a scattering process which generates correlations in the 
two-particle Schr\"odinger wave function.
Continuous variable quantum systems are 
also an interesting topic in quantum information theory
(cf.~the recent review \cite{braunloo}).
For instance, continuous versions of quantum teleportation 
\cite{braukimb} and quantum cryptography \cite{hille} can be found.

In particular we will be dealing with Gaussian states, 
since properties of these
states can be often obtained in an analytic fashion such that 
there is also an extensive theoretical literature 
about these states \cite{ferrarolect,simsud}. 
Gaussian wave functions are also a quite natural assumption.
We will investigate how the amount of generated entanglement depends on the 
ratio of the widths of the incoming particle wave functions.

In quantum optics the entanglement production for photons 
can be realized experimentally and
has been extensively studied, e.~g.~for
squeezed states hitting a beam splitter \cite{silberh,kimbuz}. In contrast to that, 
there are only a few studies of entanglement production for scattering of 
two particles \cite{tal,law}. This is surprising as scattering theory itself is a major topic 
in quantum mechanics. In \cite{law} entanglement production for scattering of two
Gaussian particles with hard core repulsion (cf.~the potential 
in eq.~(\ref{defpot})) was studied, and the present paper can be understood as 
a generalization of this paper which yields additional physical understanding. For, in \cite{law} only
the special case was treated, when the two particles have equal masses and equal initial widths of their wave 
functions. It was then found that only transient entanglement, i.~e.~non--vanishing entanglement 
during collision, can be produced. For a more general repulsive potential only a rather small quantity of 
asymptotic (permanent) entanglement could be obtained with the initial conditions of \cite{law}. In \cite{brieg} 
one can find calculations of entanglement in spin gases, where the position coordinates are treated classically. 

Our calculation looks quite similar to the calculation of
the entanglement that is generated in a beam splitter when at least
one input state
is squeezed. However, the decisive difference is that
the 
action of the beam splitter is a {\it rotation} that mixes the quadrature amplitudes of different modes, whereas the scattering process 
is a {\it reflection} with respect to the non-orthogonal 
coordinate system given by the relative coordinates. 

We now outline the organization of this paper. In section \ref{secmod} we state the scattering process and review the 
exact solution of the time--dependent Schr\"odinger equation. In section \ref{secentang} we calculate the asymptotic entanglement 
of the two particles after the scattering. Here we use extensively the covariance matrix of Gaussian states.  
A geometric interpretation of the results about entanglement is given in section \ref{secwave} where we study the
wave function of the two particles after scattering. Therewith we can also understand in detail under which conditions significant 
entanglement is produced -- and also when no entanglement at all is generated. In section \ref{secconc} we sum up the obtained results and 
point out some possibilities for further research.
\section{The scattering process and its exact solution} \label{secmod}
Here we consider the quantum mechanical scattering process of two particles in one spatial dimension 
which interact via the potential $V (x_1 -x_2 ) $. As the  scattering potential shall model hard--core repulsion, it
is given by
\bequ
V \leb |x_1 - x_2 | \rib := \left\{ \begin{array}{c @{\; \; \;} c} 0 \; ,  & |x_1 - x_2 | > a \\
\infty \; , & |x_1 - x_2 | \leq a \end{array} \right. 
\label{defpot}
\eequ
In contrast to \cite{law} we consider general mass ratios of the two particles. Hence the 
Hamiltonian for the corresponding one dimensional scattering process is
\bequ
H = \frac{\hat{p}_1^2}{2 \, m_1} + \frac{\hat{p}_2^2}{2 \, m_2} +  V \leb |x_1 - x_2 | \rib
= \frac{\hat{p}_s^2}{2 \, M_s } + \frac{\hat{p}_r^2}{2 \, M_r } +  V \leb | x_r | \rib \quad ,
\label{defham}
\eequ
where we have defined the momenta
\bea
\hat{p}_s & := & \hat{p}_1 + \hat{p}_2 \; , \nonumber \\
\hat{p}_r & := & \mu_2 \; \hat{p}_1 - \mu_1 \; \hat{p}_2 
\label{newmom}
\eea
and center of mass and relative coordinate $x_s$ and $x_r $, respectively, 
\bea
x_s & = & \mu_1 \; x_1 + \mu_2 \; x_2 \; , \nonumber \\ 
x_r & = & x_1 - x_2 \quad ,
\label{coochan}
\eea
using the mass fractions
\bequ
\mu_1 := m_1 / (m_1 + m_2) \; , \quad  \; \mu_2 := m_2 / (m_1 + m_2) \quad . 
\eequ
Besides, in eq.~(\ref{defham}) we use the masses
$ M_s:= (m_1 + m_2) /2 $ and $ M_r:= (m_1 \; m_2 ) 
/ (m_1 + m_2) $. We thus see that the Hamiltonian (\ref{defham}) decouples in the coordinates $(x_s, \, x_r)$.

Our starting configuration at $t = 0$ are two Gaussian wave packets with widths $ \sigma_1^2 $ and 
$ \sigma_2^2 $. Let the mean value of particle $1$ be located at $Q_1 \gg a$, that of 
particle 2  at $ - Q_2 \ll -a $. The momentum of the first particle 1 shall 
be $ - K $ and that of the second particle $ K $. 
We first define a state $ | f_0 \rangle \; $ with wave function  
\bequ 
f_0 ( x_1 , \, x_2 ) = \phi_G \leb x_1 \, ; Q_1, \, -K, \, \sigma_1^2 \rib \; \, 
\phi_G \leb x_2 \, ; \, - Q_2, \, K, \, \sigma_2^2 \rib \; , 
\label{initcond}
\eequ
where the Gaussian (located at $Q$ and with momentum $K$) is defined as
\bea 
\phi_G \leb  \, ; \, Q, \, K, \, \sigma^2 \rib :=  \alpha ( \sigma^2 ) \; \exp \leb i \, K \, x \rib \;
\exp \Big( - \frac{(x-Q)^2}{2 \, \sigma^2 } \Big) \; , \; \; \alpha ( \sigma^2 ) :=  
\frac{1}{\sqrt{\sigma} \; \pi^{1/4} } \; \; .
\label{defgauss}
\eea   
Note that $|f_0\rangle$ itself does not define  
a physical initial condition for any finite $Q$, since the Gaussians 
always overlap. In a more formal approach the 
two wave packets have to be 
starting from an ``infinite distance''\footnote{Following standard methods of scattering theory, 
we have to transform the state $|f_0\rangle$ 
backwards in time according to the free evolution $\exp(-iH_0t)$ 
that is obtained from $H$ by dropping the $V$-term.
Then we obtain a well-defined unitary transformation for the scattering process
by the following limit. Apply the backwards directed free evolution
$\exp(i\, H_0\, t)$, apply then the interacting time evolution $\exp(-2 \, i \, H \, t)$  and transform
the result again backwards in time by $\exp(i \, H_0 \, t)$. The $S$ matrix is then obtained in the limit
$t\to \infty$.}. 

Now we would like  to find a solution 
$ | \psi_t \rangle \in {\cal H} $ of the time--dependent 
Schr\"odinger equation 
with the Hamiltonian $H$ in eq.~(\ref{defham})
that coincides for $t \to -\infty$ approximatively with
$|f_t\rangle:=\exp(-iH_0t)|f_0\rangle$, where
\bequ
H_0 = \frac{p_s^2}{2 \, M_s } + \frac{p_r^2}{2 \, M_r } \quad . 
\label{freeham}
\eequ
is the free Hamiltonian.
In the relative coordinate 
$ x_r $ we have to satisfy the following (boundary) conditions: 
\begin{description}
\item[(A)] For  all $x_r \in [-a , \, a ] $ we have
 $ \psi_t( x_s, \, x_r)=0  \;,\;\;\; \forall t\in \R $\,.
\item[(B)] The solution $ \psi_t( x_s, \, x_r) $ is continuous in $x_r$ at the boundary of the 
potential $ x_r = a$.
\item[(C)] For $x_r \not \in  [-a , \, a ] $ the solution $ \psi_t( x_r, \, x_s) $ obeys the {\it free}
Schr\"odinger equation with Hamiltonian $H_0$.
\end{description}

We now construct the solution as follows. 
%First let us 
%evolve the initial state $ | f_0 \rangle $ with the free Hamiltonian $ H_0 $
%\bequ
%| f_t \rangle : = \exp \leb - i \, H_0 \; t \rib \; | f_0 \rangle \quad .
%\eequ
For an arbitrary wave function $ \phi (x_r)$ we introduce the unitary  
operator $ P_a $ that reflects the relative coordinate at $x_r = a$, i.~e.
\bequ
\leb P_a \, | \phi \rangle \rib (x_r) := \phi (2a - x_r) \quad .
\eequ

It can be easily seen that the operator $ \openone \otimes P_a $ commutes with the free Hamiltonian 
(\ref{freeham})
\bequ
[ \openone \otimes P_a , \; H_0 ] = 0 \quad .
\label{commrel}
\eequ
Here it is important to keep in mind that the tensor product structure
refers
to relative coordinates and not to the particle Hilbert spaces.  
Let us define the state
\bequ
| g_t \rangle : = (\openone \otimes P_a ) \; | f_t \rangle  \quad \,.
\label{defcorres}
\eequ
Because of the commutation relation (\ref{commrel}) 
the evolution $t \mapsto  | g_t \rangle $ is also a solution of the free
Schr\"odinger equation. Now the difference $ | f_t \rangle -  | g_t \rangle $ 
satisfies condition (B). In order to enforce condition (A) we use a Heaviside 
function, and thus the general solution that we were looking for 
can be written as
\bequ
\psi_t (x_s , \, x_r ) = \leb \; f_t (x_s, \, x_r)  
- g_t (x_s, \, x_r) \; \rib \; \theta (x_r -a ) \;. 
\label{gensol}
\eequ
Since we have 
\bequ
\lim_{t\to-\infty} \|\,|\psi_{t}\rangle - |f_t\rangle\|=0
\eequ
and
\bequ
\lim_{t\to\infty} \|\,|\psi_{t}\rangle - |g_t\rangle\|=0
\label{asympgt}
\eequ
we consider the mapping $|f_t\rangle \mapsto |g_t\rangle$ 
as the scattering process. 

It is important to note that due to the hard--core repulsion particle 1 always stays to
the right of particles 2. In other words, there is only reflection and no transmission for this particular potential.
\section{Entanglement between the two particles after scattering} \label{secentang}
In order to calculate the entanglement between the particles 
after the scattering, we have to consider 
solution (\ref{gensol}) for times $ t \to \infty $. Since eq.~(\ref{asympgt}) holds and the free evolution
$\exp(-iH_0t)$ does not change the entanglement, we have
\bequ
\lim_{t\to \infty} E(|\psi_t\rangle)=E(|g_0\rangle) \; ,
\eequ
where $ |g_0\rangle =  (\openone \otimes P_a ) \; | f_0 \rangle $ (cf.~eq.~(\ref{defcorres})).
It is very important to note that the state $ | g_0 \rangle$ is a 
Gaussian state \cite{ferrarolect}\footnote{A Gaussian 
state has a Wigner function that is a multi--dimensional Gaussian function.} 
For, the initial state $  | f_0 \rangle $ is obviously Gaussian and the
reflection in relative coordinates 
does not change the Gaussian nature of the state. In order to obtain the 
entanglement of $ | g_0 \rangle$, we associate the following
reduced density operator to the state $ | g_0 \rangle $ 
\bequ
\rho^{\rm red} := {\rm Tr}_2 \leb  | g_0 \rangle \, \langle  g_0 | \rib 
\label{reddeg0}
\eequ
The reduced density operator of a Gaussian state is also Gaussian.

The entanglement of a Gaussian state can be determined from its covariance matrix
that we now introduce. The four standard canonical operators of our two particle system are
\bequ
\mathbf{R} := ( \hat{x}_1 , \,  \hat{p}_1, \, \hat{x}_2 , \, \hat{p}_2 )^T \quad .
\label{standoper}
\eequ
The 4 by 4 covariance matrix of a density operator $ \rho $ is then given as
\bequ
\sigma_{i \, j} = \frac{1}{2} \, {\rm Tr} \leb \rho \; \{ R_i , \, R_j \} \rib - {\rm Tr} \leb \rho \; R_i \rib
\;  {\rm Tr} \leb \rho \; R_j \rib \; , \quad i, \, j \in \{1, \, \ldots , \, 4 \} \; ,
\eequ
where $ \{ \, \, , \, \, \} $ denotes the anticommutator.
We recall the following fact (cf.~e.~g.~\cite{ferrarolect}): a linear change of canonical operators 
with a 4 by 4 symplectic matrix $F$ and a displacement $\mathbf{D}$ , i.~e.
\bequ
\tilde{\mathbf{R}} = F \; \mathbf{R} + \mathbf{D} \; , 
\eequ
implies for the covariance matrix $ \tilde{\sigma }$ of the new canonical operators $ \tilde{\mathbf{R}}$ 
\bequ
\tilde{\sigma} = F \; \sigma \; F^{T} \quad .
\label{transsig}
\eequ 
In order to obtain the covariance matrix of the state 
\bea
\rho^\prime := | g_0 \rangle \, \langle  g_0 | = U \; \rho \; U^\dagger 
\quad {\rm with} \; \; \rho :=  | f_0 \rangle \, \langle  f_0 | \; , \quad 
U := \openone \otimes P_a \quad ,
\label{statege}
\eea
we start with the covariance matrix for $\rho = | f_0 \rangle \, \langle  f_0 | $ and the standard operator
coordinates $\mathbf{R}$ of eq.~(\ref{standoper})
\bea
\sigma := {\rm Tr} \leb \mathbf{R} \,  \mathbf{R}^T \, \rho \rib 
- {\rm Tr} \leb \mathbf{R} \, \rho \rib \; {\rm Tr} \leb  \mathbf{R}^T \, \rho \rib 
   =
\left( \begin{array}{c c c c} 
\sigma_1^2/2 & 0 & 0 & 0 \\
0  &   1/(2 \; \sigma^2_1 ) &  0 &  0  \\
0  &  0  &  \sigma_2^2/2 & 0 \\
0  &  0  &  0  &   1/ (2 \; \sigma^2_2 )
\end{array} \right) \quad ,
\label{matrsig}
\eea
as can be easily calculated from formulas (\ref{initcond}) and (\ref{defgauss}).
The covariance matrix for $ \rho^\prime $ in eq.~(\ref{statege}) in standard canonical 
coordinates $\mathbf{R}$ can be transformed as (using eq.~(\ref{statege}))
\bea
\sigma^\prime_{i \, j} & = &  {\rm Tr} \leb \rho^\prime \; R_i \; R_j \rib 
- {\rm Tr} \leb \rho^\prime \; R_i \rib \; {\rm Tr} \leb \rho^\prime \; R_j \rib
\nonumber \\
& = &  {\rm Tr} \leb \rho \; (U^\dagger \, R_i \, U) \;  (U^\dagger \, R_j \, U) \rib 
-  {\rm Tr} \leb \rho \; (U^\dagger \, R_i \, U) \rib \;  {\rm Tr} \leb \rho \; 
(U^\dagger \, R_j \, U) \rib \quad .
\label{umfpri}
\eea

We now seek a symplectic transformation which relates the new operators $ U^\dagger \, \mathbf{R} \, U $
to the standard operators $ \mathbf{R}$. Then we can use formula (\ref{transsig}) and obtain $ \sigma^\prime $
from $ \sigma $ in (\ref{matrsig}).
Since the operator $ U = \openone \otimes P_a $ acts on center of mass and relative coordinate, we
define the vector $ \mathbf{S} $ of canonical operators in this frame as
\bequ
\mathbf{S} =  ( \hat{x}_s , \,  \hat{p}_s, \, \hat{x}_r , \, \hat{p}_r )^T \quad .
\eequ
As can be easily seen from eqs.~(\ref{coochan}) and (\ref{newmom}), we have
\bequ
\mathbf{S} = F \; \mathbf{R} \quad {\rm with} \quad F = \left( 
\begin{array}{c c c c} 
\mu_1 & 0 & \mu_2 & 0 \\
0  &  1  &  0 &  1  \\
1  &  0  &  -1 & 0 \\
0  &  \mu_2  &  0  &  - \mu_1
\end{array} \right) \quad .
\eequ
The identity
\bequ
U^\dagger \, \mathbf{R} \, U =  F^{-1} \, U^\dagger \, \mathbf{S} \, U 
\label{anwuas}
\eequ
follows immediately.
As $U$ is a reflection in the relative coordinate, we can easily calculate
\bequ
U^\dagger \, \mathbf{S} \, U = G \, \mathbf{S} + \mathbf{D} \quad  {\rm with} \; \;
G = \left( 
\begin{array}{c c c c} 
1 & 0 &  0 & 0 \\
0  &  1  &  0 &  0  \\
0  &  0  &  -1 & 0 \\
0  &  0  &  0  &  - 1
\end{array} \right) \; \; , \; 
\mathbf{D} = \left( \begin{array}{c} 0 \\ 0 \\ 2\, a \\ 0 \end{array} \right) \; .
\label{intromag}
\eequ
Using eq.~(\ref{anwuas}) and (\ref{intromag}), we arrive at
\bequ
U^\dagger \, \mathbf{R} \, U =  F^{-1} \, G \, F \,  \mathbf{R} +   F^{-1} \, \mathbf{D} \quad .
\eequ
Thus the covariance matrix $ \sigma^\prime $ is related to the covariance matrix $ \sigma $ 
of eq.~(\ref{matrsig}) by
\bequ
\sigma^\prime = F^{-1} \, G \, F \; \sigma \; F^T \, G^T \, \leb F^{-1} \rib^T \quad .
\label{prodmatr}
\eequ 
We write $ \sigma' $ in the form 
\bequ
\sigma^\prime = \leb \begin{array}{c c} A & C \\ C^T & B \end{array}  \rib \; ,
\label{formcov}
\eequ 
where $A, \; B $ and $ C $ are 2 by 2 matrices.
Then a straightforward computation of the product of matrices in eq.~(\ref{prodmatr}) yields
\bea
A & = & \leb   \begin{array}{c c}  2 \, \mu_2^2 \, \sigma_2^2 + \frac{ (\Delta \mu)^2 \, \sigma_1^2 }{2}  &  0 \\
0  & \frac{ 2 \, \mu_1^2 }{\sigma_2^2} + \frac{ (\Delta \mu)^2 }{2 \, \sigma_1^2 } \ \end{array}  \rib \; , \quad
B = \leb   \begin{array}{c c}  2 \, \mu_1^2 \, \sigma_1^2 + \frac{ (\Delta \mu)^2 \, \sigma_2^2 }{2}  &  0 \\
0  & \frac{ 2 \, \mu_2^2 }{\sigma_1^2} + \frac{ (\Delta \mu)^2 }{2 \, \sigma_2^2 } \ \end{array}  \rib \; ,
\nonumber \\
C & = &  \leb   \begin{array}{c c} \Delta \mu \,
\leb \mu_1 \, \sigma_1^2 - \mu_2 \, \sigma_2^2 \rib   &  0 \\
0  & \Delta \mu \, \leb \frac{ \mu_2^2 }{\sigma_1^2} - \frac{ \mu_1}{\sigma_2^2} \rib \end{array} \rib \quad \,,
\label{resultcov}
\eea
where we used the abbreviation
\bequ
\Delta \mu = \mu_1 - \mu_2 \quad .
\eequ

Looking at the definitions of covariance matrix and reduced density operator, it is clear that the covariance 
matrix which corresponds the reduced state $ \rho^{\rm red} $ of eq.~(\ref{reddeg0}) is given by the submatrix $A$
of eq.~(\ref{resultcov}). It is known 
\cite{agarwal,seraopt,botero} that the von 
Neumann entropy -- as 
defined in eq.~(\ref{basedefi}) -- of a one mode Gaussian state $ \rho^{\rm red} $ is
\bequ
S \leb  \rho^{\rm red} \rib = \leb d + \frac{1}{2} \rib \; \log_2  \leb d + \frac{1}{2} \rib 
- \leb d - \frac{1}{2} \rib \; \log_2  \leb d - \frac{1}{2} \rib \quad ,
\label{entandd}
\eequ 
where the quantity $d \geq 1/2$ is given as
\bequ
 d^2 = \det(A) = \det (B) = 4 \, \mu_1^2 \, \mu_2^2 + \leb \Delta \mu \rib^2 \, \left[ 
\frac{ \leb \Delta \mu \rib^2 }{4}
+ \frac{ \mu_1^2 \, \sigma_1^2 }{\sigma_2^2} + \frac{ \mu_2^2 \, \sigma_2^2 }{\sigma_1^2} \right] \quad .
\label{resuld2}
\eequ
If we insert the (positive) 
value of $ d (\mu_1, \, \mu_2, \, \sigma_1^2 , \, \sigma_2^2 ) $ into eq.~(\ref{entandd}),
we get the desired asymptotic entanglement in terms of the four parameters
$ \mu_1, \, \mu_2, \, \sigma_1^2 , \, \sigma_2^2 $.

Instead of calculating the von Neumann entropy of the reduced density operator one can also consider
-- as done in \cite{law} -- the purity of the state $ \rho^{\rm red} $ 
\bequ
{\cal P} \leb \rho^{\rm red} \rib := {\rm Tr} \leb   \left[ \rho^{\rm red} \right]^2 \rib \quad .
\eequ
The smaller $ {\cal P} \leb \rho^{\rm red} \rib $, the larger is the generated entanglement. It is 
known \cite{ferrarolect} that the purity of a Gaussian state like $  \rho^{\rm red} $ is given by
\bequ
{\cal P} \leb \rho^{\rm red} \rib = \frac{1}{2 \; \sqrt{ \det(A) } } = \frac{1}{2 \, d} \quad.
\eequ
Thus the purity is -- as the von Neumann entropy - characterized by the determinant of the covariance matrix
$A$. 

We now discuss our result for the entanglement as expressed in eqs.~(\ref{entandd}) and (\ref{resuld2}). 
We first note that $d^2$ depends only on the two width ratios $ \sigma_1^2 /  \sigma_2^2 $,  $ \sigma_2^2 
/  \sigma_1^2 $ and on the mass fractions $ \mu_1 $, $ \mu_2 $ of the two particles. We see that for 
a large $d$ - and therfore a large entanglement -- it is necessary that one of the two width ratios is large. 
In contrast to that the quantities $ 0 < \mu_1, \, \mu_2 < 1$ enter in formula (\ref{resuld2}) not as a 
quotient, so that choosing large mass ratios does not generate a large $d$. We will discuss the conditions for the
production of a lot of entanglement in detail in section \ref{secwave}.

One can easily check that always $ d \geq 1/2 $ and thus $ S \leb  \rho^{\rm red} \rib \geq 0 $. 
If $ d = 1 / 2$, then the entanglement vanishes according 
to eq.~(\ref{entandd}). No entanglement is generated for the two conditions
(cf.~eq.~(\ref{resuld2}))
\begin{enumerate} 
\item $ \mu_1 = \mu_2 = 1/2 $, i.~e.~the two scattered particles have equal mass. This result was already obtained
in \cite{law}.
\item $ \mu_1 \, \sigma_1^2 = \mu_2 \, \sigma_2^2 $, i.~e.~the heavy particle has a small width,
the light particle has a large width.
\end{enumerate} 
\section{The wave function after scattering}    \label{secwave}
In this section we study the wave function $ g_0 \leb x_1, \, x_2 \rib $ that corresponds to the 
state $ |  g_0 \rangle $ \footnote{Note that this in not the asymptotic wave 
function after the scattering. However -- as explained before -- the entanglement of the two 
particles can just as well be obtained from the state $ |  g_0 \rangle $.}. 
This will give us more insight into the results in eqs.~(\ref{entandd}) and~(\ref{resuld2}) for 
the generated entanglement. In complete analogy to the calculation of the covariance matrix  
$ \sigma^\prime $ in the last section, we can obtain the wave function of the state $ |  g_0 \rangle
= (\openone \otimes P_a ) \; | f_0 \rangle $ in three steps from the initial wave function $ f_0 ( x_1 , 
\, x_2 ) $ of eq.~(\ref{initcond}): 
\begin{enumerate}
\item Replace in $ f_0 ( x_1 , \, x_2 ) $ the coordinates $(x_1, \, x_2 )$ with the 
coordinates $(x_s, \, x_r)$ using the inverse to relation
(\ref{coochan}).
\item Apply the operator $ ( \openone \otimes P_a ) $ by changing the variable $ x_r $ to $ ( 2 \, a - x_r ) $.
\item Go back to the coordinates $ (x_1, \, x_2 )$ with relation (\ref{coochan}).
\end{enumerate}
The result of this procedure is
\bea
g_0 ( x_1 , \, x_2 ) & = & \phi_G \leb 2 \, \mu_2 \, x_2 + (\mu_1 - \mu_2) \, x_1  ; \, Q_1 - 2 \, \mu_2 \, a , 
\, -K, \, \sigma_1^2 \rib \; \nonumber \\
& & \times \; \phi_G \leb   2 \, \mu_1 \, x_1 - (\mu_1 - \mu_2) \, x_2 ; 
\, - Q_2 + 2 \, \mu_1 \, a , \, K, \, \sigma_2^2 \rib \; , 
\label{rohg0}
\eea
where the Gaussians $ \phi_G(\; ) $ have been defined in eq.~(\ref{defgauss}). Since the entanglement of
a Gaussian state depends only on its covariance matrix, any displacement of the wave function in position or momentum 
space does not change its entanglement. Thus instead of $ g_0 ( x_1 , \, x_2 ) $ in eq.~(\ref{rohg0}) we introduce
the simpler wave function $ \tilde{g} (x_1, \, x_2 ) $ that keeps only the relevant quadratic terms in 
$ x_1, \, x_2 $ in the exponent 
\bea
\tilde{g}_0 ( x_1 , \, x_2 ) & = & \alpha \leb \sigma_1^2 \rib \; \exp \Big( - \frac{ [ (\mu_1 - \mu_2) \, x_1 
+  2 \, \mu_2 \, x_2 ]^2}{2 \, \sigma_1^2 } \Big) \nonumber \\
& & \times \; \alpha \leb \sigma_2^2 \rib \; \exp \Big( - 
\frac{ [ 2 \, \mu_1 \, x_1 - (\mu_1 - \mu_2) \, x_2 ]^2}{2 \, \sigma_2^2 } \Big) \; .
\label{simpliwave}
\eea
This wave function can be rewritten with a quadratic form in the exponent
\bequ
\tilde{g}_0 ( x_1 , \, x_2 ) = \alpha \leb \sigma_1^2 \rib \; \alpha \leb \sigma_2^2 \rib \; 
\exp \Big( - \frac{1}{2} \, \mathbf{x}^T \, L^T \, \Sigma \, L \, \mathbf{x} \Big) \; , 
\label{resuess}
\eequ
where we have introduced $  \mathbf{x}^T := (x_1, \, x_2) $ and the two matrices 
\bequ
L :=  \leb \begin{array}{c c} \mu_1 - \mu_2 & 2 \, \mu_2 \\  2 \, \mu_1  & \mu_2 - \mu_1 \end{array}  \rib \; , \quad 
\Sigma :=   \leb \begin{array}{c c} 1/\sigma_1^2 &  0 \\  0 &   1/\sigma_2^2 \end{array}  \rib \quad .
\label{defmartl}
\eequ 
Defining the matrix in the exponent of the wave function in eq.~(\ref{resuess}) as
\bequ
M :=  L^T \, \Sigma \, L  \quad , 
\label{defmarmm}
\eequ
a comparison with eq.~(\ref{simpliwave}) yields for its entries
\bea
M_{1 \, 1} & := & \frac{(\mu_1 - \mu_2)^2}{\sigma_1^2} + \frac{4 \, \mu_1^2}{\sigma_2^2}\; , \; \; 
M_{2 \, 2} := \frac{4 \, \mu_2^2}{\sigma_1^2} + \frac{(\mu_1 - \mu_2)^2 }{\sigma_2^2} \; ,  \nonumber \\
M_{1 \, 2} & = & M_{2 \, 1} := 2 \, (\mu_1 - \mu_2) \; 
\leb \frac{\mu_2}{\sigma_1^2} - \frac{\mu_1}{\sigma_2^2} \rib \quad .
\label{defmmat}
\eea
>From eq.~(\ref{resuess}) we see that the wave function 
factorizes into a product of particle 1 and 2 wave functions, 
if and only if $  M_{1 \, 2} = 0$. Eq.~(\ref{defmmat}) shows us 
that this is the case when $ \mu_1 = \mu_2 $ or when
$ \mu_1 \, \sigma_1^2 = \mu_2 \, \sigma_2^2 $ -- in accordance with our findings in the last section. 
% A physics argument why these conditions lead to no entanglement will be given in section.

We now discuss the two cases with factorizing wave functions in more detail. In the case $\mu_1 = \mu_2$ the
widths $ \sigma_1^2 , \, \sigma_2^2 $ of the particles are interchanged after the scattering process, as the 
formulas for $M _{1 \, 1} $ and $ M_{2 \, 2} $ in eq.~(\ref{defmmat}) show\footnote{Here we can neglect the 
fact that Gaussian wave functions broaden in free time evolution, as the scattering process normally takes place on
a short time scale compared to this effect.}. More generally, it can be shown that the two particles exchange
the shape of their wave functions in the scattering process, if the two particle masses are equal.
When $ \mu_1 \, \sigma_1^2 = \mu_2 \, \sigma_2^2 $, a short calculation shows that 
\bequ
M_{1 \, 1} = \frac{1}{\sigma_1^2} \; , \quad M_{2 \, 2} = \frac{1}{\sigma_2^2} \; ,
\eequ
and thus both particles keep their initial width. That no entanglement occurs for $ \mu_1 \, \sigma_1^2
= \mu_2 \, \sigma_2^2 $ depends crucially on the Gaussian shape of the incoming particle wavefunction. For,  
the two mixed terms proportional to $x_1 \, x_2$ in the two exponents of eq.~(\ref{simpliwave})
have opposite sign, if $ \mu_1 \, \sigma_1^2 = \mu_2 \, \sigma_2^2 $, and therefore cancel each other.  
\bef
\begin{center}
\mbox{\epsfig{figure=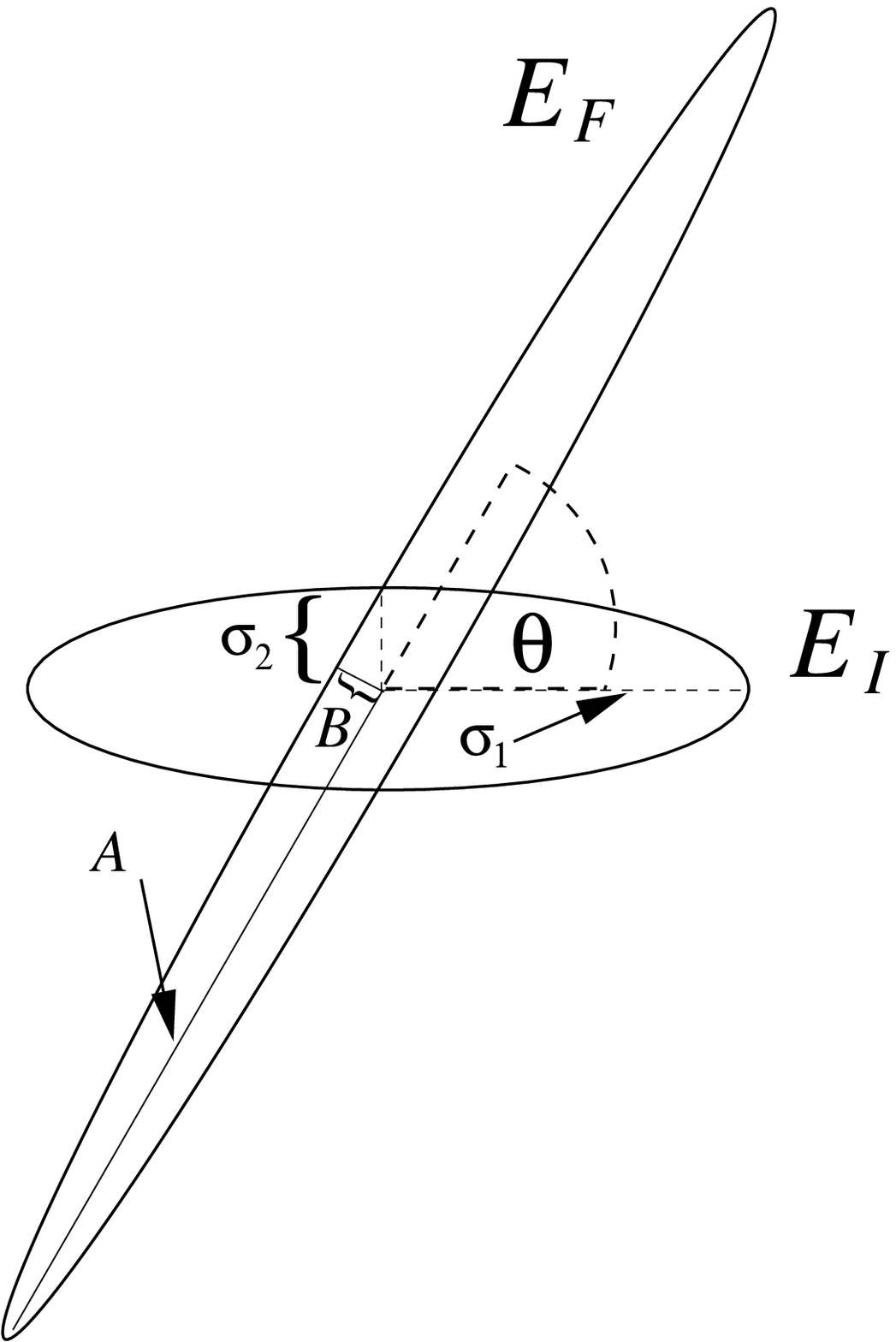, width=80mm}}
\end{center}
\caption{Here the initial ellipse $E_I$ with semiaxes $ \sigma_1, \, \sigma_2$ and 
the final ellipse $E_F$ with semiaxes $ A, \, B$ are depicted. The rotation angle 
$ \theta $ of $E_F$ is approximately $ 63,4^0$ and thus the length of 
the semimajor axis $A$ is more than doubled in comparison to $\sigma_1$, whereas 
the semiminor axis $B$ is shrunk by the same factor in comparison to $ \sigma_2 $.}
\label{elli}
\eef

Points $(x_1, \, x_2)$ where the wave function $ \tilde{g}_0 ( x_1 , \, x_2 ) $ 
has the same value form an ellipse. For example, we may select  
those points $(x_1, \, x_2)$ where the exponent in eq.~(\ref{resuess})  
is $ -1/2 $. This leads to an analytical expression for an ellipse $E_F$ as follows
\bequ
E_F: \quad \quad \mathbf{x}^T \, M \,  \mathbf{x} = 1 \quad .
\label{defelli}
\eequ
Generally this ellipse is oblique to the $x_1, \, x_2$ axes; this is precisely then the case when
entanglement occurs. The semimajor axis $A$ and the semiminor axis $B$ of the ellipse $E_F$ in 
eq.~(\ref{defelli}) are determined from the two eigenvalues $ \lambda_1, \,  \lambda_2 $ of $M$ 
\bequ
A = \frac{1}{\sqrt{\lambda_2}} \; , \quad B = \frac{1}{\sqrt{\lambda_1}} \; , 
\quad (\lambda_1 \geq \lambda_2) \quad .
\label{majorel}
\eequ
Let $ \theta $ be the angle between the semimajor axis of $E_F$ and the $x_1$ axis. $ \theta $ is determined by the
eigenvectors of $M$. In fig.~\ref{elli} we depict the geometric quantities $A$, \, $B$ and $ \theta $. We can also 
associate to the initial wave function 
$ f_0 ( x_1 , \, x_2 ) $ in eq.~(\ref{initcond}) an ellipse by keeping -- as done for the wave
function $  \tilde{g}_0 ( x_1 , \, x_2 ) $ -- only the quadratic terms in the exponent yielding
\bequ
\tilde{f}_0 ( x_1 , \, x_2 ) =   \alpha \leb \sigma_1^2 \rib \; \alpha \leb \sigma_2^2 \rib \; 
\exp \Big( - \frac{1}{2} \, \mathbf{x}^T \, \Sigma \, \mathbf{x} \Big) \; , 
\label{initellps}
\eequ
where the diagonal matrix $ \Sigma $ has been defined in eq.~(\ref{defmartl}). 
The quadratic form in the exponent of equation (\ref{initellps}) defines 
an ellipse $E_I$ where the major axes are the widths $ \sigma_1, \,  \sigma_2 $ of the 
two particles. We adopt the convention that $ \sigma_1^2 \geq  \sigma_2^2 $. Thus we can understand
the effect of scattering in a geometric way as a transformation of the initial ellipse $E_I$ 
into the final ellipse $E_F$ (see fig.~\ref{elli}). Because of eq.~(\ref{defmarmm}) and $ \det L = -1$,  
\bequ
\det M = \det \Sigma  = \frac{1}{\sigma_1^2 \, \sigma_2^2} \; ,
\eequ  
and thus the areas of the ellipses $E_I$ and $E_F$ are equal. 

In general we get for the major axes of the final ellipse (cf.~eq.~(\ref{majorel})) rather 
complicated expressions in our parameters $ \mu_1, \, \mu_2, \, \sigma_1^2 , \, \sigma_2^2 $. 
A significant simplification takes place, if we consider the case that one 
particle has a much greater width than the other one, i.~e. the semimajor axis 
of the initial ellipse $E_I$ is much longer than the semiminor axis ($ \sigma_1 \gg  \sigma_2 $).  
This is also the most interesting case, as only then a lot of entanglement can be created via scattering, as
we have remarked before. When $ \sigma_1 \gg  \sigma_2 $, it holds approximately that the semimajor axis of the 
ellipse $E_I$ is mapped via the transformation $ L $ in eq.~(\ref{defmartl}) to the semimajor axis of the ellipse $E_F$.
This means that
\bequ
\mathbf{P} := L \; \left( \begin{array}{c} \sigma_1 \\  0 \end{array} \right) =  \sigma_1 \; 
\left( \begin{array}{c} 2 \, \mu_1 -1  \\  2 \, \mu_1 \end{array} \right) \quad ,
\label{vectorp}
\eequ
and the length $A$ of the semimajor axis in $E_F$ is 
\bequ
A \approx \| \mathbf{P} \| =  \sqrt{Q(\mu_1)} \; \sigma_1 \; ,
\label{resuaxis}
\eequ
where we define the quadratic polynomial 
\bequ
Q(x) := 8 \, x^2 - 4 \, x + 1 \quad .
\label{defpoly}
\eequ
Since the areas of the ellipse $E_I$ and $E_F$ are equal, we get for the length of the semiminor axis
\bequ
B \approx \frac{ \sigma_2 }{ \sqrt{Q(\mu_1)}} \; .
\label{resuaxis2}
\eequ
In the same approximation ($ \sigma_1 \gg  \sigma_2 $) the angle $  \theta $ of the semimajor axis (cf.~fig~\ref{elli}) can 
be read off from the vector $\mathbf{P}$ in eq.~(\ref{vectorp}) as \footnote{In general this angle depends also on the values $ \sigma_1^2 , 
\, \sigma_2^2 $ of the initial ellipse.}
\bequ
\theta = \arctan \leb \frac{2 \, \mu_1}{2 \, \mu_1 - 1} \rib \quad .
\label{resuthet}
\eequ
If we analyze these results, we see that for $ 0 < \mu_1 < 1$ the 
polynomial $Q(\mu_1)$ in 
eq.~(\ref{defpoly}) has values between the minimum $ 1/2$ (for $ \mu_1 = 1/4 $ ) 
and 5 (for $ \mu_1 \approx 1 $). Thus according to eqs.~(\ref{resuaxis}) and (\ref{resuaxis2})
the semimajor axis $A$ can be elongated by up to a factor $ \sqrt{5} $ or shortened by up to a factor $ \sqrt{2} $. 
The semiminor axis is always scaled by the inverse factor. Looking at the angle $ \theta $ in eq.~(\ref{resuthet}),
we find 
%  \arctan(2) entspricht 63,4 Grad 
\bea
\mu_1 > \mu_2  \quad & \Rightarrow & \quad \arctan(2) < \theta < \frac{\pi}{2} \; , \nonumber \\
\mu_1 < \mu_2  \quad & \Rightarrow & \quad \frac{\pi}{2}  < \theta < \pi \quad .
\eea
If $ \mu_1 = \mu_2 = 1/2$ we have $ \theta = \pi / 2$ which is
consistent with the generation of no entanglement. If $ \mu_1 \to 0 $ we get 
$ \theta = \pi $ from eq.~(\ref{resuthet}), and there is no entanglement as well. In our approximation 
the case $ \mu_1 \to 0 $ corresponds to the case $ \mu_1 \, \sigma_1^2 =  \mu_2 \, 
\sigma_2^2 $ studied above, since terms of the magnitude $ \sigma_2^2 / \sigma_1^2 $ have been neglected.
In general, our analysis shows that the scattering process leads to more complicated transformations 
of the ellipse than in the situation of squeezed states hitting a beamsplitter where the initial ellipse 
is only rotated (cf.~\cite{kimbuz}). 
\bef
\begin{center}
\mbox{\epsfig{figure=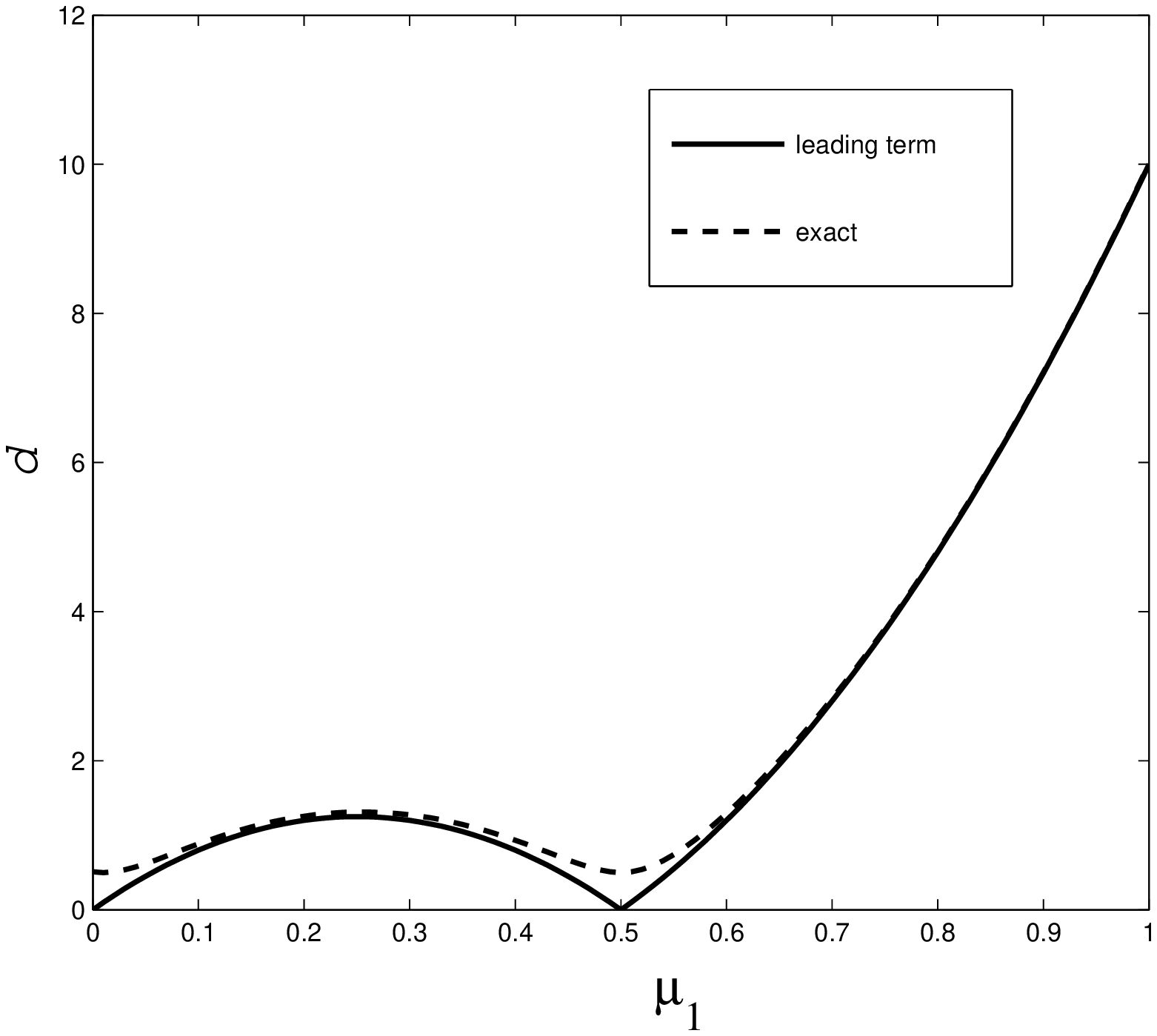, width=120mm}}
\end{center}
\caption{Here the quantity $d$ is plotted as a function of $\mu_1$ for fixed width ratio  $ \sigma_1 / \sigma_2 = 10 $.
The solid line shows the leading term in formula (\ref{asymtd}), the dahed line the exact value in eq.~(\ref{resuld2}).}
\label{enta}
\eef

How can we understand the circumstances the production of a lot of entanglement in this geometric picture\,? 
First, for $ \sigma_1 \gg  \sigma_2 $ we calculate the leading term in $  \sigma_1 / \sigma_2 $ of the quantity $d$ 
from eq.~(\ref{resuld2}) 
\bequ 
d \sim | 2 \, \mu_1 - 1  | \; \mu_1 \; \frac{ \sigma_1 }{ \sigma_2 } \quad .
\label{asymtd}
\eequ
For the fixed value of $ \sigma_1 / \sigma_2 = 10 $ we plot in fig.~\ref{enta} 
the exact $d$ and this approximation as a functions of $\mu_1 $ and find very good agreement. 
Thus we will use the simpler form of eq.~(\ref{asymtd}) for our subsequent discussion.

According to fig.~\ref{enta} (sold line) we find a local maximum of 
$d$ at $ \mu_1 = 1/4$ and the absolute maximum at $ \mu_1 = 1 $. The value at
the absolute maximum is greater by a factor 8 than that at the local one. If one looks
only at the angle of the final ellipse $E_F$, one would expect the absolute maximum for $ \mu_1 = 1/4$, since for
this value we get the angle $ \theta = 3 \, \pi / 4$ from  eq.~(\ref{resuthet}). 
Here the ellipse $E_F$ is more tilted in the $x_1, \, x_2 $ plane than at the value $  \mu_1 = 1 $ 
where we obtain $ \theta = \arctan(2) \approx  63,4^0$. The explanation why the value of
$d$ for $ \mu_1 = 1 $ is greater than that for $ \mu_1 = 1/4 $ 
lies in the factor $ \sqrt{Q(\mu_1)} $ in eqs.~(\ref{resuaxis}) and (\ref{resuaxis2})
that can lead to a stretching or shrinking of the initial ellipse $E_I$ . 
For, at $ \mu_1 = 1/4 $ the polynomial $ Q(\mu_1) $ has -- as mentioned before -- its minimum with 
value $1/2$ and thus 
\bequ 
\frac{A(\mu_1 = 1/4 )}{B(\mu_1 = 1/4 )} = Q(\mu_1 = 1/4) \; \frac{\sigma_1}{\sigma_2} 
= \frac{1}{2} \; \frac{\sigma_1}{\sigma_2} \quad .
\eequ
In contrast to that, for $ \mu_1 = 1 $ we get
\bequ 
\frac{A(\mu_1 = 1)}{B(\mu_1 = 1 )} = Q(\mu_1 = 1) \; \frac{\sigma_1}{\sigma_2} 
= 5 \; \frac{\sigma_1}{\sigma_2} \quad .
\eequ
Thus the ratio between semimajor and semiminor axis is 
augmented 
 by a factor 5, 
whereas for $ \mu_1 = 1/4 $ it is diminished  by a factor 2.
This effect outweighs the effect of the angle $ \theta $ and explains why the absolute maximum is located at
$ \mu_1 = 1 $. 
\section{Conclusions} \label{secconc}
We have studied in detail the scattering of two Gaussian wave packets with hard core repulsion. The generated 
entanglement could be calculated analytically (cf.~our central result in eqs.~(\ref{entandd}) and~(\ref{resuld2})\,).
The special hard--core potential that we have considered yields Gaussians as asymptotic wave functions. Therefore, 
we could apply results about Gaussian states. 

As we vary for the two particles the ratio of their masses 
and the ratio of their initial widths, it turns out that a great width ratio
is necessary for the production of much entanglement in the scattering. If we vary for fixed and large width ratio
the mass ratio of the two particles, maxima of entanglement production are reached,  
\begin{enumerate}
\item If the mass of the particle with large width is much greater than that of the particle with small width. This is the absolute maximum.
\item If the mass of the particle with large width is approximately $1/3$ of the mass of the particle with small width (local maximum 
as shown in fig.~\ref{enta}).
\end{enumerate}
By contrast, no entanglement is generated, 
\begin{enumerate}
\item If the two particles have equal mass. 
\item If for masses and widths of the two particles the relation $ \mu_1 \, \sigma_1^2 = \mu_2 \, \sigma_2^2 $ holds, i.~e.~
one particle has large width and small mass, the other particle small width and large mass. 
\end{enumerate}
These results about maxima and minima of entanglement are in our opinion not at all obvious, even though scattering at a 
hard core potential is one of the simplest scattering processes. 

We have gained additional geometric understanding in our results by looking at the wave function after scattering. This Gaussian 
wave function defines an ellipse whose properties determine the degree of entanglement after scattering. This shows strong
parallels to studies of entanglement in quantum optics where an ellipse can be defined for squeezed light hitting a beam splitter.
However in contrast to the beamsplitter, the scattering process can not only rotate the ellipse but also stretch or squeeze it which is
relevant for the amount of entanglement.

Our results indicate that there remains much to be done in the subject of entanglement production via scattering. 
Beyond the special case of hard core repulsion one should look at more realistic interaction potentials which allow
for reflection and transmission of the particles. Also the modification of our results for bosons and fermions poses an 
interesting problem. We leave these tasks to future studies.
\section*{Acknowledgment}
The authors are grateful to the Landesstiftung Baden--W\"urttemberg that supported
this work in the program ''Quantum-Information Highway A8'' (project ''Kontinuierliche Modelle
der Quanteninformationsverarbeitung'').
%
% \bibliography{/.mnt/boleyn/users1/schmues/paper1/bibdokt}

%
\end{document}